\documentclass[10pt,letterpaper,twocolumn]{article}
\usepackage[latin1]{inputenc}
\usepackage{amsmath}
\usepackage{amsfonts}
\usepackage{amssymb}
\usepackage[dvips]{graphicx}
\author{A. A. Arellano-Baeza(a), R. V. Garcia(b), and M. Trejo-Soto(b)
\\
\\ (a) Mining Department, University of Santiago de Chile 
\\ (b) Earth Sciences School, Autonomous University of Sinaloa}
\title{Study of the structural changes in the Popocatepetl volcano in Mexico related to microseismicity by applying the lineament analysis to the Aster (Terra) satellite data.}
\begin{document}
\maketitle

\begin{abstract}
Mexico is one of the most volcanically active regions in North America. Volcanic activity in central Mexico is associated with the subduction of the Cocos and Rivera plates beneath the North American plate. Periods of enhanced microseismic activity, associated with the volcanic activity of the Popocatepetl volcano is compared with periods, during which the microseismic activity was low. We detected systematical changes in the number of lineaments, associated with the microseismic activity due to lineament analysis of a temporal sequence of high resolution satellite images of the Popocatepetl volcano, provided by the ASTER/VNIR instrument. The Lineament Extraction and Stripes Statistic Analysis (LESSA) software package was used for the lineament extraction. In the future it would allow develop a methodology for detection of possible elevation of pressure in volcano edifice. 
\end{abstract}

\section{Introduction}
Popocatepetl is the most famous volcano in Mexico located about 70 km to the southeast of Mexico City (Long. 261.37, Lat. 19.07, elevation 5465 m). It is one of several active volcanoes that form the Trans-Volcanic Belt of Mexico (also know as Neo-Volcanic Axes) and its existence is related to the geodynamics of the North American and Coco plates. It has an almost symmetric cone that rises 4,200 m above the surrounding plain. The volcano is composed of two distinct structures: the old part, called Nexpayantla, and the newer structure, which stands on the southearstern part of the old Popocatepetl. The crater is 400 by 600 m wide with vertical walls. Prevalent volcanic activities, observed in the volcano, are Plinian, Vulcanian, pheatomagmatic, lava flows and fumaroles. The most recent eruption began in December 1994.

   Similarly to other volcanoes, the Popocatepetl volcano seismicity is represented by of micro- or small earthquakes in form of swarms, the magnitude of which is generally below 4.0. Presence of earthquake swarms is normally related to volcanic activity even if the surface eruptions are not observed. Moreover, the migration of earthquakes usually suggests the potential movement of magma toward the surface. However, only a few studies have been made to establish a relationship between the appearance and migration of hypocenters and the volcanic processes of intrusions. In particular Gambino et al., (2004) analyzed the location of microseismicity preceding the 2002-2003 Mt. Etna eruption. They found that seismicity and swarms occurring since February 2002 near the central crater can be interpreted as a response of the volcano to magma movements along the NNW-SSE volcano-genetic trend. In particular, the intense micro-seismic activity observed between August 31 and September 3, 2002 beneath the SSE Rift and eastward from it may be related to rising of magma.	

   Earthquake swarms and ground deformation are amongst the most frequently cited precursors of volcanic eruptions (Scarpa and Tilling, 1996). In particular, Voight (1988, 1989) suggested a forecasting model based on dynamics of material failure to describe accelerating earthquake event rate and ground deformation. It is based on the assumption that the growth of magmatic pathways is driven by rock failure in the volcanic edifice under sustained, near-constant fluid pressure in the magma chamber. Therefore the rate of magma ascent is limited by the rate of fracture growth, recorded as seismicity. Recently Chastin and Main (2003) have tested the Voigts hypothesis using the seismic data from volcano Kilauea, Hawaii. They have found that acceleration of earthquake rates is not observed systematically before all individual eruptions.  It can be related to the fact that there are a number of processes leading to the accumulation of strain and therefore to the ground deformation. When the initial period of volcano-wide deformation is generally attributed to the inflation of the volcanic edifice by the rising magma, the thermal expansion and faulting/fracturing can also contribute to the strain accumulation (expansion) o relaxation (fracturing) (Jackson et al., 1998).

   During the last decade, the application of Synthetic Aperture Radar (SAR) interferometry data for ground deformation studies on volcanoes allowed to generate deformation time series which significantly contribute to better understanding of its sources (see, for example, Schmidt and Burgmann, (2003), Pritchard and Simons, (2002)). Remy et al (2007) studied the inflation of the Aira Caldera (Japan) using the SAR interferometry ERS data during a three year period. They estimate a roughly constant annual rate of inflation of about m3/year that corresponds to the ground deformation of about 2 cm, which agrees with JERS and GPS measurements.

    SAR interferometry technique has also been successfully applied to study of ground deformations associated with earthquakes related to the movement of tectonic plates (see for example (Satybala, 2006; Schmidt and Brgmann, 2006, Lasserre et al., 2005, Funning et al., 2005)).  Observed deformations were of the same order of magnitude as in case of volcano deformation. Recently Arellano et al. (2004, 2006, 2007) have found that presence of earthquake can strongly affect the system of lineaments extracted from the high-resolution ASTER satellite images. It was found that a significant number of lineaments appears approximately one month before strong earthquake. One month after the earthquake the lineament configuration returns to its initial state. These features were detected during the analysis of 8 strong earthquakes in South and North America and China. These features were not observed in tests areas, situated far from any earthquake. Singh V.P and Singh R.P. (2005) used the lineament analysis to study changes in stress pattern around the epicenter of the Mw=7.6 Bhuj earthquake occurred 26 January 2001 in India. The results obtained also confirm that the lineaments retrieved from the images 22 days before the earthquake differ from the lineaments obtained 3 days after the earthquake. It was assumed that they are related to fractures and faults and that their orientation and density give an idea about the fracture pattern of rocks. The results also show the high level of correlation between the continued horizontal maximum compressive stress deduced from the lineament and the earthquake focal mechanism.

    The main question is how the lineaments extracted from 15-30 m resolution images are able to reflect the accumulation of stress deep in the earth crust when the ground deformation, associated with earthquakes is about a few centimeters? The term lineaments, originally proposed by Hobbs (1904), defined lineaments as significant lines of landscape that reveal the hidden architecture of the rock basement. However O'Leary et. al. (1976) defined lineament as a simple or composite linear feature of a surface whose parts are aligned in a rectilinear or slightly curvilinear relationship and which differ from the pattern of adjacent features and presumably reflects some sub-surface phenomenon. It supposes that lineaments are able to detect, at least partially, the presence of ruptures deep in the Earths crust. This means that corresponding image processing of extraction of system of lineaments using different algorithms (see, for example, Karnieli et al., 1996, Fitton and Fox, 1998, Wang et al., 1990, Zlatopolsky, 1992, 1997) makes possible to retrieve (at least partially) the information about the dynamics of the Earths crust kilometers and even tens of kilometers underground that is disseminated over a surface. Deeper events are disseminated over bigger areas and require lower resolution images for their detection. That is why 10-30 m resolution images are useful for earthquake studies. Although they are not able to see some individual crack, they are able to integrate (at least) the information about the presence of faults tens of kilometers underground, and track the changes related to accumulation or relaxation of strength due to the movement of tectonic plates. 

   In this paper we will use the same methodology to detect possible changes in the lineament structure related to the microseismicity events related to the volcanic activity of the Popocatepetl volcano.
    
\section{Instrumentation and Data Analysis}
For this study we used the images from the Advanced Spaceborne Thermal Emission and Reflection Radiometer (ASTER) on board of the TERRA satellite. The satellite was launched to a circular solar-synchronous orbit with altitude of 705 km. The radiometer is composed by three instruments: Visible and Near Infrared Radiometer (VNIR) with 15 m resolution (bands 1-3), Short Wave Infrared Radiometer (SWIR) with 30 m resolution (bands 4-9) and Thermal Infrared Radiometer TIR with 90 m resolution (bands 11-14) which measure the reflected and emitted radiation of the Earths surface covering the range 0.56 to 11.3 $\mu$m (Abrams, 2000). 

    The images were processed using the Lineament Extraction and Stripes Statistic Analysis (LESSA) software package (Zlatopolsky, 1992, 1997), which provides a statistical description of the position and orientation of short linear structures through detection of small linear features (stripes) and calculation of descriptors that characterize the spatial distribution of stripes. The program also makes it possible to extract the lineaments  straight lines crossing a significant part of the image. To make this extraction, a set of very long and very narrow (a few pixels) windows (bands), crossing the entire image in different directions, was used. In each band the density of stripes, the direction of which is coincident with the direction of the band, is calculated. When the density of stripes overcomes a pre-established threshold, the chain of stripes along the band is considered as a lineament. The value of threshold depends on the brightness of the image, relief, etc. and is established empirically. Previous studies showed that lineaments, extracted from the image by applying the LESSA program, are strongly related to the main lineaments, obtained from the geomorphological studies (Zlatopolsky, 1992, 1997). The details about the application of LESSA package for earthquake studies are given in (Arellano et al., 2006).

We studied a time sequence of images covering the Popocatepetl volcano area during 2000-2005. Unfortunately the level of cloudiness was high, and from 10 images analysed we selected only four of them covering exactly the same area with 0 percents of clouds (see Figure 1 first line). The area extracted from the VNIR band 1Level 2 images correspond to the north-east part of the volcano and have the coordinates, upper left corner: $19^{\circ}$07'56''N, $98^{\circ}$37'20''W; upper right corner: $19^{\circ}$07'56''N, 983007W; lower right corner: $19^{\circ}$00'55''N, $98^{\circ}$30'07''W; lower left corner:$19^{\circ}$00'55''N, $98^{\circ}$37'20''W. The corresponding systems of lineaments were extracted by applying the LESSA program, using the same parameters (Figure 1, second line). As it can be seen, there are similarities in the structure of lineaments, nevertheless their density changes significantly from image to image, such that we ordered the image sequence from high to low lineament density. Subsequent lines contain the information about the depth, longitude, and latitude of microseismic events 30 days before (grey) and 90 days after (black) of the time of each image. The last line represents the three-dimensional view of the same events. These data were provided by CENAPRED (http://www.cenapred.unam.mx/es/). 
\onecolumn
\begin{figure}
\includegraphics[width=18cm]{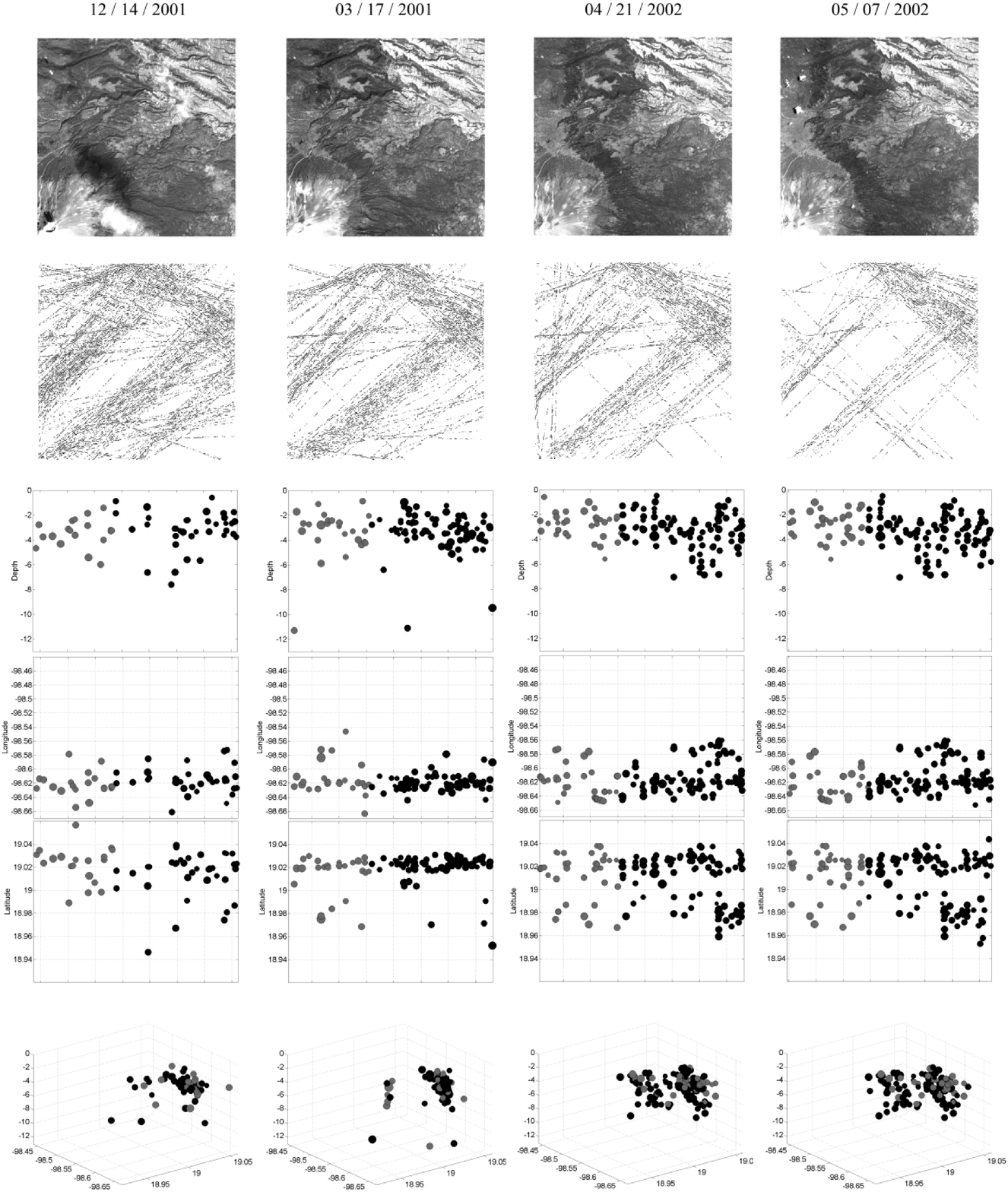} 

\caption{ From top to bottom: fragments of ASTER/VNIR image of the Popocatepetl volcano (band 1), systems of lineaments, extracted from the images above, depth, latitude, and longitude of micro earthquakes 30 days before (gray) and 90 days after (black) to the time of each image, three-dimensional distribution of micro earthquakes 30 days before (gray) and 90 days after (black).}
\end{figure} 
\twocolumn
\begin{figure}[t]
\includegraphics[width=8.5cm]{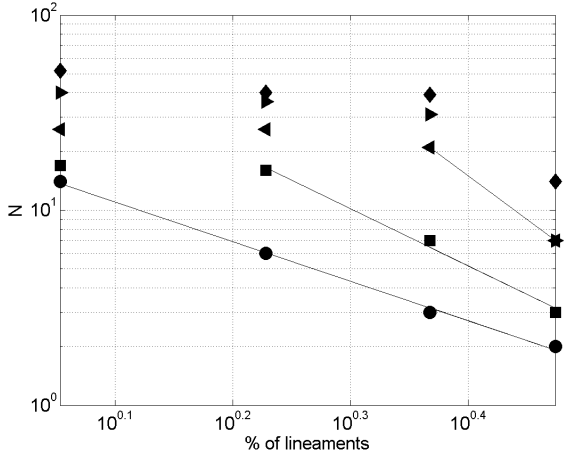} 

\caption{ Relationship between the density of lineaments and the number of earthquakes during next 10, 20, 30, 40 and 50 days. }
\end{figure} 

As it can be seen, there is a relationship between the lineament density and the number of earthquakes. In order to quantify this relationship at least in very elemental way, we plot a relationship between the percentage of pixels corresponding to the lineaments and the number of earthquakes during the next 10, 20, 30, 40 and 50 days (Figure 2). To our surprise, a very interesting inverse power low dependence with an exponent -2.030.07 appeared between the percentage of pixels and the number of earthquakes ten days ahead. After that this relationship starts to saturate gradually. This is rather surprising and has to the best of our never been described. We do not have at this point any explanation to offer leaving these results as a prospective goal to be investigated.
    
\section{Discussion and conclusions}
Analysis of a relationship between the structure of lineaments and microseismicity observed in the Popocatepetl volcano, showed a power low relationship between the percentage of pixels corresponding of lineaments extracted from the ASTER satellite image and the number of earthquakes ten days ahead. This result is opposite to that obtained previously for earthquakes not related to the volcano activity. In that case the number of lineaments increased significantly a few weeks before an earthquake (Arellano et al. 2004, 2006, 2007). This discrepancy can be explained assuming that in the last case the main reason of earthquakes is compression and accumulation of strength in the Earths crust due to subduction of tectonic plates, whereas in the first case we deal with the inflation of a volcano edifice due to elevation of pressure and magma intrusion. So it is natural to expect that the lineament behavior would be different. 
   This study represents a first step in understanding of how the processes of volcanic seismic activity are reflected in the lineament structure. It is necessary to increase significantly the number of events analysed and develop more sophisticated methodology including both more precise quantification of lineament behavior, and more precise analysis of seismic events, including their location, energy release and waveform.
   Nevertheless, the results obtained are relevant because open the possibility of volcano monitoring using satellite images. It is important because to date only a very low percentage of volcanoes is constantly monitored using very efficient but expensive ground chains. The use of new techniques would allow in the future to detect possible elevation of pressure in the volcano edifice that makes it possible to concentrate on a special attention to this specific volcano and to initiate other studies using more precise but unfortunately much more expensive satellite and ground based techniques.

\textbf{Acknowledgments}

We acknowledge Hiroji Tsu (Geological Survey of Japan CSJ)  ASTER Team Leader, Anne Kahle (Jet Propulsion Laboratory  JPL)  US ASTER Team Leader and the Land Processes Distributed Active Archive Center for providing the ASTER level 2 images. We acknowledge A. Zlatopolsky for providing the Lineament Extraction and Stripes Statistic Analysis (LESSA) software package and helpful suggestions. We acknowledge the CENAPRED Mexico for providing the Popocatepetl volcano seismic data. This work has been supported by DICYT/USACH grant.
\\
\\
\textbf{REFERENCES}

Abrams, M., The Advanced Spaceborne Thermal Emission and Reflection Radiometer (ASTER): Data products for the high spatial resolution imager on NASA's Terra platform, International Journal of Remote Sensing 21(5), 847-859, 2000.

Arellano-Baeza, A.A., Zverev, A., Malinnikov, V. Study of the structure changes caused by earthquakes in Chile applying the lineament analysis to the Aster (Terra) satellite data (Abstract). 35th COSPAR Scientific Assembly, Paris, France, July 18-25, 2004.

Arellano-Baeza, A.A., Zverev, A., Malinnikov, V. Study of the structure changes caused by earthquakes in South America applying the lineament analysis to the Aster (Terra) satellite data, Advances in Space Research  37(4), 690-697, doi:10.1016/j.asr.2005.07.068, 2006.

Arellano-Baeza, A.A., Garcia, R.V., Trejo-Soto, Use of high resolution satellite images for tracking of changes in the lineament structure, caused by earthquakes, submitted to Advance in Space Research, 2007.

Gambino S., Mostaccio, A.,  Patane, D., Scrafi, L., Ursino A. High-precision locations of the microseismicity preceding the 2002-2003 Mt. Etna eruption. Geophys. Res. Lett., 31, L18604, doi:10.1029/2004GL020499, 2004.

Chastin, S. F. M., Main, I. G. Statistical analysis of daily seismic event rate as a precursor to volcanic eruptions. Geophys. Res. Lett., 30, 13, 1671, doi:10.1029/2003GL016900, 2003.

Funning, G.J., Parsons, B., Wright, T.J.,  Jackson, J.A., Fielding, E.J.  Surface displacements and source parameters of the 2003 Bam (Iran) earthquake from Envisat advanced synthetic aperture radar imagery, Surface displacements and source parameters of the 2003 Bam (Iran) earthquake from Envisat advanced synthetic aperture radar imagery, 110(B9), doi: 10.1029/2004JB003338, 2005. 

Jackson, P., Shepherd J. B., Robertson, R.E.A., Skerritt, J. Ground deformation studies at Soufriere Hills Volcano, Montserrat I: Electronic distance meter studies. Geophys. Res. Lett., 25(18), 3409-3412, 1998.

Fitton, N.C., Cox, S.J.D. Optimizing the application of the Hough Transform for the automatic feature extraction from geoscientific images. Computers and Geosciences, Vol. 24, P. 933-951, 1998.

Hobbs, W.H. Lineaments of the Atlantic border region, Geological Society American Bulletin, 15, 483-506, 1904.

Lasserre, C., Peltzer, G., Cramp� F., Klinger, Y., Van der Woerd, J., Tapponnier, P. Coseismic deformation of the 2001 Mw = 7.8 Kokoxili earthquake in Tibet, measured by synthetic aperture radar interferometry, J. Geoph. Res., 110(B12), doi: 10.1029/2004JB003500, 2005.

O'Leary, D.W., Friedman, J.D., Pohn, H.A. Lineament, linear lineation some proposed new standards for old terms. Geological Society America Bulletin, 87, 1463-1469., 1976.

Karnieli, A., Meisels, A., Fisher, L., Arkin, Y.Automatic Extraction and Evaluation of Geological Linear Features from Digital Remote Sensing Data Using a Hough Transform. Photogrammetric Engineering and Remote Sensing. 62(5), 525-531, 1996.

Pritchard, M.E., Simons, M. A satellite geodetic survey oflarge scale deformation of volcanic centres in the central Andes, Nature, 418, 167170, 2002.

Remy, D., Bonvalot, S., Murakami, M., Briole, P., Okuyama, S. Inflation of the Aira Caldera (Japan) detected over Kokubu urban area using SAR interferometry ERS data, eEarth, Volume 2, Issue 1, 2007, pp.17-25.

Satyabala, S.P. Coseismic ground deformation due to an intraplate earthquake using synthetic aperture radar interferometry: The Mw6.1 Killari, India, earthquake of 29 September 1993, J. Geoph. Res., 111(B2), doi: 10.1029/2004JB003434, 2006.

Scarpa, R., Tilling, R.I. (Eds), Monitoring and mitigation of volcano hazards, 841 pp., Springer-verlag, Berlin, 1996.

Singh, V.P., and R.P. Singh, Changes in stress pattern around epicentral region of Bhuj earthquake, Geoph. Res. Lett., 32, doi: 10.1029/2005GL023912, 2005.

Schmidt, D.A., Burgmann, R. Time-dependant land uplift andsubsidence in the Santa Clara valley, California, from a large interferometric synthetic aperture radar data set, J. Geophys. Res., 108, 2416, doi:10.1029/2202JB002267, 2003.

Schmidt, D.A., Brgmann, R. InSAR constraints on the source parameters of the 2001 Bhuj earthquake, Geoph. Res. Lett., 33(2), doi: 10.1029/2005GL025109, 2006.

Voight, B. A method for prediction of volcanic eruptions, Nature, 332, 10, 125130, 1988. 

Voight, B. A relation to describe rate-dependent material failure, Science, 243, 200203, 1989. 

Wang, J., Howarth, P.J. Use of the Hough Transform in Automated Lineament Detection. IEEE Tran. Geoscience and Remote Sensing, Vol. 28. No. 4. P. 561-566, 1990.

Zlatopolsky, A.A. Program LESSA (Lineament Extraction and Stripe Statistical Analysis): automated linear image features analysis  experimental results, Computers and Geosciences, 18(9), 1121-1126, 1992.

Zlatopolsky, A.A. Description of texture orientation in remote sensing data using computer program LESSA, Computers and Geosciences, 23(1), 45-62, 1997.

\end{document}